\documentclass[page-classic,linenumbers]{epl2} %for one column style with line numbers
\usepackage{amsfonts}
\usepackage{amsmath}
\usepackage{amssymb}
\usepackage{graphicx}
\usepackage[square,sort&compress,comma,numbers]{natbib}%
\usepackage{caption}
\usepackage{ulem}
\usepackage[labelsep=quad,indention=10pt]{subfig}
\title{Bundle formation in  parallel aligned polymers with competing interactions}
\author{Sandipan Dutta\inst{1} \and P. Benetatos\inst{2}\thanks{E-mail: \email{pben@knu.ac.kr}} 
\and Y.S. Jho\inst{1,3}\thanks{E-mail: \email{ysjho@apctp.org}} }
\institute{                    
  \inst{1} Asia Pacific Center for Theoretical Physics, Pohang, Gyeongbuk, 790-784, Korea\\
  \inst{2} Department of Physics, Kyungpook National University, 80 Daehakro, Bukgu, Daegu 702-701, Korea \\
  \inst{3} Department of Physics, Pohang University of Science and Technology, 790-784, Korea
}

\abstract{
Aggregation of like-charged polymers is widely observed in biological and soft matter systems. 
In many systems, bundles are formed when a short-range attraction of diverse physical origin like charge-bridging, hydrogen-bonding or
hydrophobic interaction, overcomes the longer-range charge repulsion. In this Letter, we present a general mechanism of
bundle formation in these systems as the breaking of the translational invariance in  parallel aligned polymers 
with competing interactions of this type. We derive a criterion for finite-sized bundle formation as
well as for macroscopic phase separation (formation of infinite bundles).
}

\begin{document}

\maketitle

\textbf{Introduction} -
In nature, like-charged polymers such as proteins, DNA and viruses, often form self-assembled structures like bundles, gels or clumps~\cite{israelachvili2011intermolecular,
ruben1997alzheimer,netz2003neutral,dobrynin2005theory,kong2009complex,Haviv_EurBioJ,li2011superlattice}. 
Like-charged bundles are ubiquitous in biological matter \cite{Lodish_book}, examples include the actin filament and microtubule bundles which are the main structural elements of the cytoskeleton.
The presence of a short-range attraction helps overcome the energetic disadvantage due to the long-range electrostatic repulsion in such systems.   
  Bundles have unique mechanical properties which are fundamentally different from those of the wormlike chain which is 
a minimal theoretical model for single semiflexible filaments  \cite{bundle_Frey}. 
%More specifically, they exhibit an effective bending rigidity which depends on the 
%bending mode and the shear modulus of the cross-links \cite{bundle_Frey}.  
Bundling of DNA is used for the storage of 
genetic information \cite{Bloomfeld_DNA}. Analyzing the formation of these supramolecular structures is important for our understanding 
of biological functions and for biomimetic applications as well. 

 The term bundling in polymer systems has been used to describe the parallel alignment of polymers into densely packed structures 
\cite{BorukhovPNAS,zilman2005role, Kierfeld_Kuhne,PB_AZ_PRL, PhysRevE.83.021905,Ha_Liu_EPL,Ha_Liu_PRL,Henle_Pincus, Claessens_Bausch,Grason_Bruinsma,Heussinger_Grason,Gov,sayar2007finite,stevens1999bundle,netz_polycations,zierenberg2015amorphous}. 
 In many studies, the bundled state is a uniform equilibrium phase and as such, in the thermodynamic limit, a bundle would have infinite lateral size.  The observed finite size in many cases has been a long-standing theoretical challenge and several explanations have been proposed, ranging from electrostatics, to filament helicity, and packing defects.
Ha and Liu \cite{Ha_Liu_EPL} have shown that bundles can be formed through correlations in charge fluctuations. This mechanism of bundle formation entails a rather high energy-barrier which limits the size of the bundles 
and traps them in metastable states. Henle and  Pincus \cite{Henle_Pincus} predict  bundles of finite size in equilibrium for large-size counterions 
or frustrated attractive interactions, but of infinite size otherwise. In a different approach which agrees with the experimental observations in F-actin
\cite{Claessens_Bausch}, Grason and collaborators \cite{Grason_Bruinsma,Heussinger_Grason} show that,
in chiral biopolymers, the inherent chirality  limits the size of the bundles.  Monte Carlo simulations by Kierfeld {\it et al.} \cite{Kierfeld_Kuhne} 
suggest that finite-size bundles may be kinetically arrested structures due to slow dynamics. In a somewhat similar spirit, Gov \cite{Gov} 
proposes that quenched packing defects (splay and twist) limit the size of the bundles in cells' cytoskeleton.

In this Letter, we approach bundle formation as microphase separation of directed polymers with competing attractive and
repulsive interactions. The idea of microphase separation originated in the field of block copolymers \cite{leibler1980theory} 
where the competing tendencies for incompatible blocks may result in the formation of regular structures.
The first proposal for the occurrence of microphase separation in polyelectrolytes is due to
the pioneering works of  Borue and Erukhimovich \cite{Boryu1988statistical}, and Joanny and Leibler \cite{joanny1990weakly}. 
The basic idea is that in weakly charged polyelectrolytes in poor solvents, there is an interplay between the tendency for polymer collapse 
(poor solvent) and the tendency of counterions to maximize their entropy. Under certain conditions, the minimization of the free energy 
results in the local violation of electroneutrality and  alternating regions of high and low polymer densities are formed
(see Ref. \cite{dobrynin2005theory} and references therein). 

 The concept of competing interactions (SALR: shorter-range attraction and longer range repulsion) is well established in the
field of colloids. 
%Groenewold and Kegel
 It has been suggested that this competition in colloids leads to the formation of finite-sized clusters 
in a fashion similar to the formation of nuclei by the interplay of strong (nuclear) and Coulomb interactions 
\cite{groenewold2001anomalously,groenewold2004colloidal}. Both cluster and gel phases in colloidal systems have been observed  experimentally and also in numerical
investigations (see Ref. \cite{Yun_Liu_ChemEng} and references therein).

In this Letter, we use the idea of competing electrostatic repulsion and shorter-range attraction of various origins,
like hydrophobic, cross-linking, or hydrogen-bonding, to study  bundle formation in like-charged biomolecular systems.
For simplicity, we assume the polyelectrolytes to be  parallel pre-aligned and investigate an instability
criteria of the free energy that leads to the formation of bundles. From this instability we deduce
a phase (stability) diagram for the formation of both finite and infinite bundles.
To illustrate the theory, we apply it to the case of like-charged polymers interacting via a Yukawa 
potential with an additional inverted Gaussian short-range attraction. We start from a semi-microscopic description of the system 
and we obtain a coarse-grained field theory in terms of an appropriate order parameter using an alternate form of the Hubbard-Stratonovich 
transformation \cite{goldbart1996randomly}, similar to the approach in Ref. 
 \cite{benetatos2013bundling}. 

\begin{figure*}[!htbp]
 \begin{center}
\includegraphics[scale=0.3,trim={{6.0cm} {6.0cm} {8.0cm} {7.0cm}},clip]{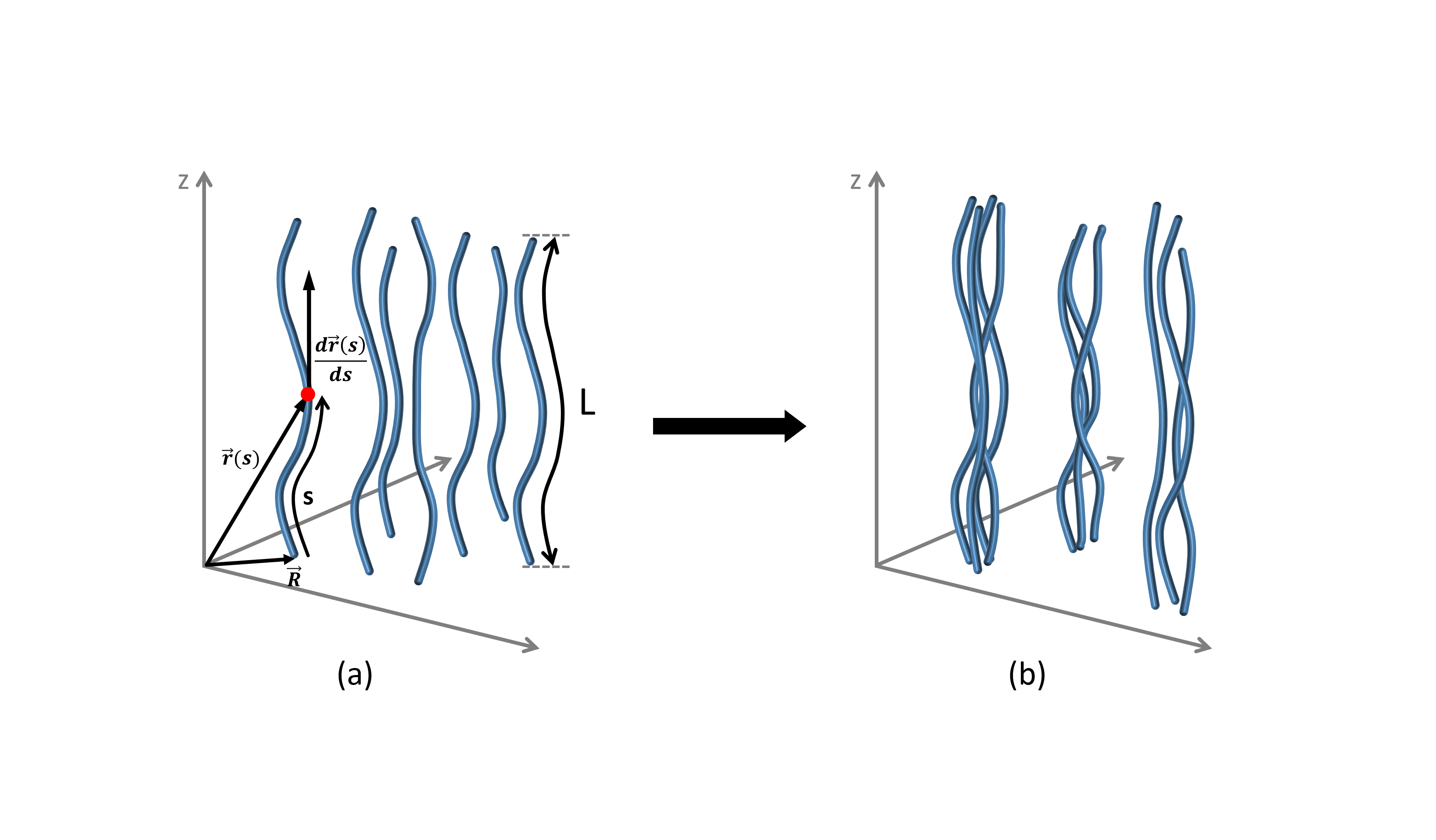}
\end{center}
\caption{A schematic diagram of charged polymers aligned along the $z$ axis. We
refer to $x$ and $y$ as the transverse or in-plane direction. Figure (a) shows the disordered in-plane (nematic liquid) state and 
Figure (b) shows the bundled state.}
\label{Fig1}
\end{figure*}

 \textbf{Model} -
 We consider a system of $N$ long and  parallel aligned uniformly charged polymers. The polymer system is assumed to be sufficiently 
 dense to be amenable to a coarse-grained mean-field treatment (i.e., in the melt regime).
The polymers have a length $L$ and a charge $q$ (thus, the line charge density is $q/L$) and have an average areal density $\rho=N/A$. We assume
the polymers are stretched along the z-axis as shown in Figure \ref{Fig1}, with their end-points freely sliding on
two parallel planes.
The electrostatic interaction in the system is described by a Debye-H{\"u}ckel (Yukawa) potential $V_{R}(\mathbf{r})\propto \exp\left({-r/\lambda_R}\right)/r$, where $\lambda_R$ is the screening length. 
We also have a short-range attractive potential $V_{A}(\mathbf{r})$ that may be a result of hydrogen-bond, hydrophobic, cation-$\pi$, charge-bridging or
reversible cross-links ~\cite{silverstein1998simple,benetatos2013bundling}. For long, weakly tilted polymers the interactions become local in the $z$ 
direction \cite{benetatos2013bundling}. The total interaction potential is denoted by $V(\mathbf{r})=V_{R}(\mathbf{r})+ V_{A}(\mathbf{r})$. 
We use the wormlike chain model to describe the polymers ~\cite{rubinstein2003polymers}. 
The single-polymer Hamiltonian has a term which expresses the bending energy of the polymer and another term which represents the parallel
alignment potential along the $z$ direction \cite{warner1985rod}
\begin{equation}
 \mathfrak{h}_{0} = \frac{\kappa}{2}\int_0^Lds\dot{\mathbf{u}}^2(s)-\kappa_{0} \int_0^L ds\left(\frac{3}{2}u_z^2(s)-\frac{1}{2}\right),
 \label{eq1}
\end{equation}
where $\kappa$ is the bending rigidity and $\kappa_{0}$ is the strength of the aligning potential. 
The latter may be due to an external interaction, e.g. with a liquid crystal matrix, or 
generated self-consistently via a Maier-Saupe potential \cite{SpakowitzWang}. However, the isotropic-nematic transition is not included in our analysis.
The $\mathbf{u}(s)=d\mathbf{r}(s)/ds$ is the tangent vector to the polymer at the point $\mathbf{r}(s)$, where $s$ is the arc-length parameter along the polymer contour. 
The Hamiltonian of the system is
\begin{equation}
\mathcal{H} = \sum_{i=1}^{N}\mathfrak{h}_{0}^{i} + \frac{1}{2L}
\int_0^Lds\sum_{i\neq j}^N V(\vert\mathbf{r}_i(s)-\mathbf{r}_j(s)\vert),
 \label{eq3}
\end{equation}
Since we consider strongly stretched polymers, these wormlike chains are transformed into a $(2+1)$-dimensional system of directed polymers \cite{panayotis2010}.
The main reason for assuming strong-stretching  is that it implies the weakly bending approximation in the polymer conformations. 
As shown in \cite{benetatos2013bundling}, the same approximation can be achieved by considering perpendicular grafting and long persistence length, with 
qualitatively similar results concerning the microphase separation. The Hamiltonian of the system in the strong stretching approximation becomes
\begin{align}
 \mathcal{H} & = \frac{\beta\epsilon}{2}\sum\limits_{i=1}^N\int_0^L dz\mathbf{u}_j^2(z)+
 \frac{1}{L}\int_0^Ldz\sum\limits_{i\neq j}\beta V(\mathbf{r}_i(z)-\mathbf{r}_j(z)), 
 \label{eq4}
\end{align}
where $u_x(z) = dx(z)/dz$ and $u_y(z)= dy(z)/dz$, and $\epsilon = 3\kappa_{0}$. We have inserted the factor $\beta = 1/k_BT$ to make the Hamiltonian dimensionless.  

\textbf{Order parameter and instability} -
The partition function can be written in terms of a field $\Omega$ after performing the Hubbard-Stratonovich
transformation of the areal density of the system (see Appendix B in Supplementary \cite{Supplementary} for details)
\begin{align}
 \mathcal{Z} & = \int\mathcal{D}{\Omega(\mathbf{k},z)}\exp(-Nf[\Omega]).
 \label{eq7}
\end{align}
The Landau-Wilson type free energy $f(\Omega)$ per polymer in the above equation is given by
\begin{align}
 f[\Omega] & = \frac{\rho}{2L}\int_0^Ldz\sum_{\mathbf{k}}\frac{1}{2}\vert \beta V(\mathbf{k})\vert\vert\Omega(\mathbf{k},z)\vert^2 -
 \frac{1}{N}\ln\mathfrak{z}, 
 \label{eq7.1}
 \end{align}
 where the single-polymer partition function $\mathfrak{z}$ is given by
 \begin{align}
 \mathfrak{z}[\Omega]
& = \biggl(\int d\mathbf{R}\mathcal{D}\mathbf{u}\exp\biggl(\int_0^Ldz\biggl(-\frac{\beta\epsilon}{2}\mathbf{u}^2(z)+\sum_{\mathbf{k}}
\vert\beta V(\mathbf{k})\vert
\Omega(\mathbf{k},z)\biggl(\Theta(\beta V(\mathbf{k})<0)
\nonumber\\&+ i\Theta(\beta V(\mathbf{k})>0)\biggr)e^{i\mathbf{k}.\mathbf{r}(z)}\biggr)\biggr)\biggr)^N.
\label{eq8}
\end{align}

Within the saddle point approximation, we get a solution for the field $\Omega(\mathbf{k},z)$ by minimizing the 
free energy in  equation \eqref{eq7.1} as
\begin{align}
 \Omega^{sp}(\mathbf{k},z) & = \left\langle \hat{\rho}^{\ast}(\mathbf{k},z)\right\rangle_{\mathfrak{z}}, 
 \label{eq9}
\end{align}
where $\rho(\mathbf{k},z)$ denotes the local areal density. The equation \eqref{eq9} is defined only in the region where the Fourier transform of the potential $V(\mathbf{k})$ is attractive, $V(\mathbf{k}) < 0$. 
The averaging $\langle..\rangle_{\mathfrak{z}}$ in  equation \eqref{eq9} is with respect to the single-polymer partition function $\mathfrak{z}[\Omega^{sp}]$.
From  equation \eqref{eq9} we see that the field $\Omega(\mathbf{k})$ acts as an order parameter of the system, capable of distinguishing between 
a liquid state in the $x-y$ plane (that is, a disordered state with uniform areal density whose Fourier modes vanish for all $\mathbf{k}\neq 0$) and a crystalline phase with Bragg peaks at the reciprocal lattice wavevectors.
The bundled phase can involve finite bundles arranged in a periodic structure or an infinite bundle  where the polymers completely phase separate into
a dilute and a more concentrated phase. In the latter case, the $k=0$ mode of the order parameter becomes unstable.

The stability of a liquid can be determined from the second-order coefficient of the expansion of the free energy in the order parameter, which in our case is $\Omega$. 
This yields the spinodal line in the same way as outlined in Ref. \cite{Fredrickson_spinodal}. If the coefficient is positive, the liquid phase is stable
and vice versa. This gives
\begin{align}
 F(\mathbf{k}^{\ast}) = \rho^{\ast}\vert\beta V(\mathbf{k}^{\ast})\vert b(k^{\ast})& > 1
 \text{ and }\beta V(\mathbf{k})<0 & \text{(unstable)} 
 \label{eq10}
\end{align}
where $b(k^{\ast}) = (-8 + 8\exp(-{k^{\ast}}^2/2) + 4{k^{\ast}}^2)/{k^{\ast}}^4$. We have defined the dimensionless wavenumber $k^{\ast} = \mathbf{k}\sqrt{L/\beta\epsilon}$
and the dimensionless areal density $\rho^{\ast} = \rho\left(L/\beta\epsilon\right)$. 
We use $\rho^{\ast} = 1$ in the rest of the discussion.
We call the function $F(\mathbf{k}^{\ast})$ the stability function. From now on, we use the notation $\mathbf{k}$ for the dimensionless
wavenumber and $\beta V(\mathbf{k})$ for the dimensionless Fourier transform of the potential. When the Fourier transform of the potential is repulsive, $\beta V(\mathbf{k})>0$, the second order 
coefficient never changes sign ( see Appendix B in Supplementary \cite{Supplementary}) and there is no instability. When the system is stable, it is in
a uniform liquid state. When it becomes unstable the order parameter becomes non-uniform and bundles are formed. As shown in Eq. (\ref{eq10}), 
for an appropriate interaction potential, the bundling instability can be triggered by lowering the temperature or by increasing the average areal density
of the system. 

\begin{figure*}[!htbp]
	\centering
	\subfloat{%
		\includegraphics[height=5.2cm]{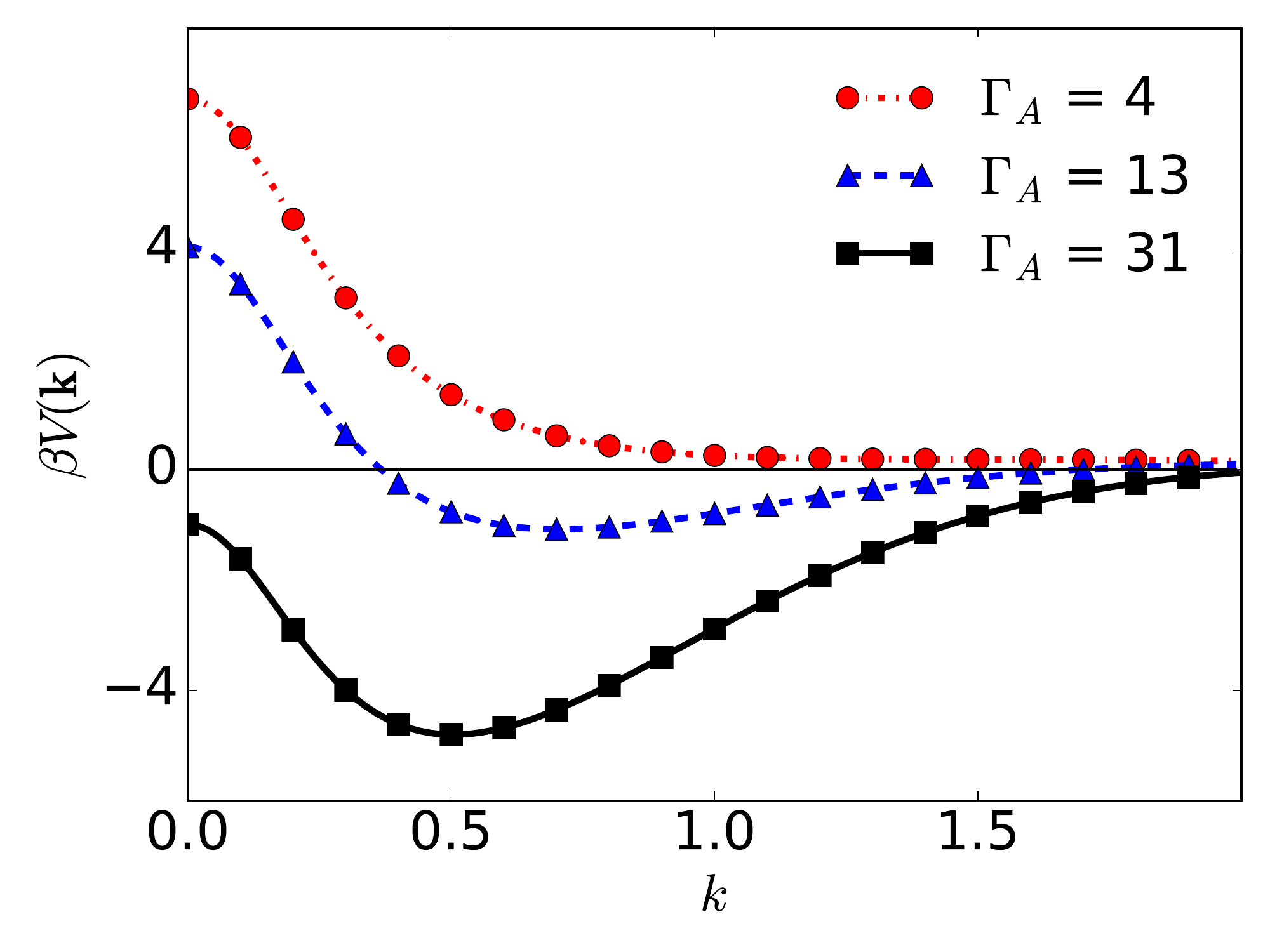}%
	}
	\subfloat{%
		\includegraphics[height=5.2cm]{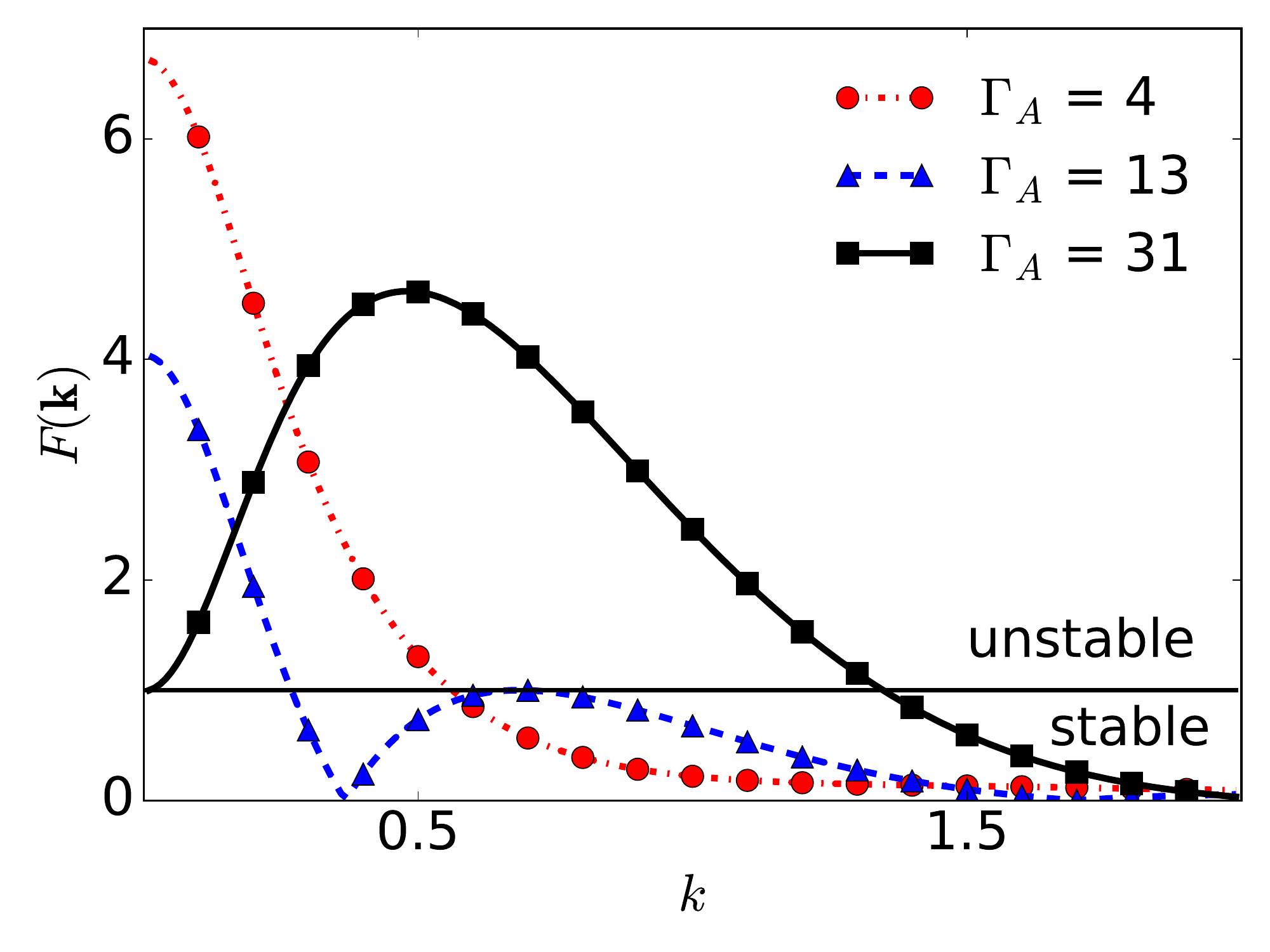}%
	}
	\caption{(a) The Fourier transform of the potential $\beta V(\mathbf{k})$ and (b) the corresponding stability function $F(\mathbf{k})$ showing the three cases,
		no bundles (circles), finite bundles (triangles) and the macroscopic phase separation (squares).
		The strength of the repulsive potential is $\Gamma_R = 5$, and the range of the repulsive and attractive 
		potentials are $\lambda_R = 0.5$ and $\lambda_A = 0.3$ respectively at $\rho^{\ast} = 1$.  $\mathbf{k}$ is dimensionless
		and is scaled with respect to the in-plane radius of gyration $\sqrt{L/\beta\epsilon}$ of the polymers.
		The $\Gamma$'s have been scaled with respect to $k_BT$ and the $\lambda$'s by the in-plane
		radius of gyration of the polymers $\sqrt{L/\beta\epsilon}$.}
	\label{Fig2}
\end{figure*}

\begin{figure*}[h]
        \centering
           \subfloat{%height=5.2cm
              \includegraphics[scale = 0.3]{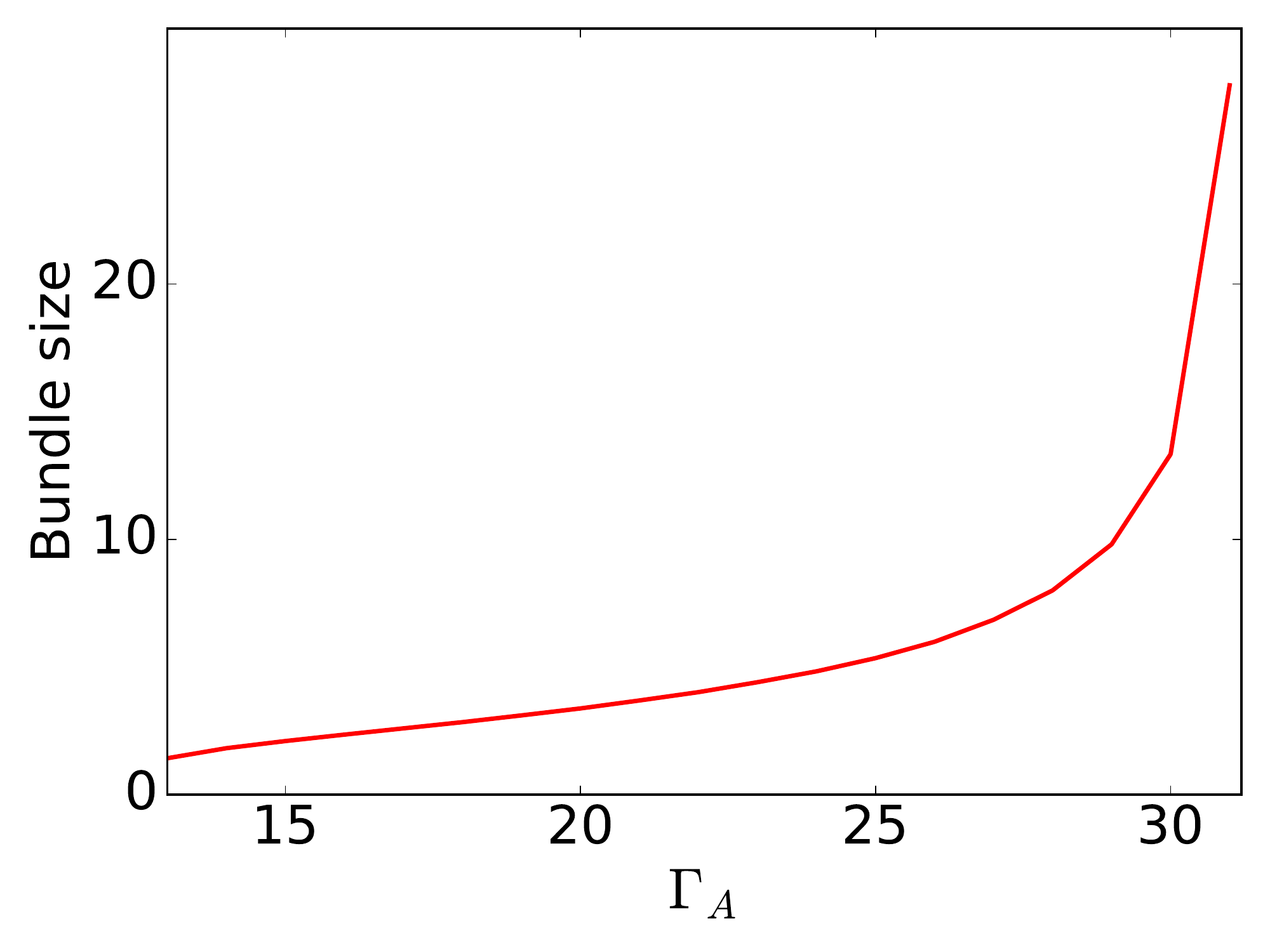}%
           }
           \subfloat{%
              \includegraphics[scale = 0.3]{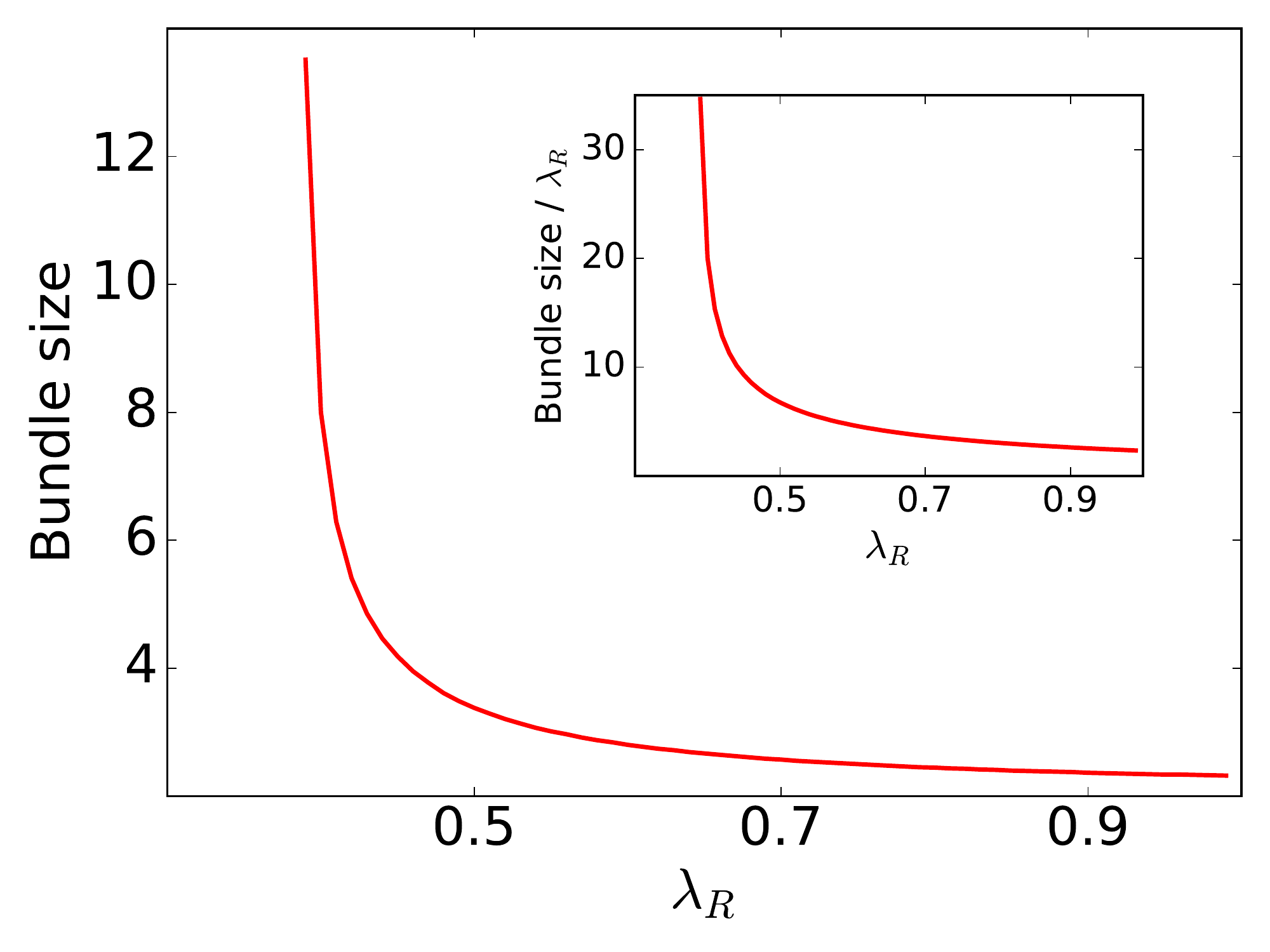}%
           }
           \caption{The size of the bundles vs (a) the strength of the attractive interactions $\Gamma_A$ at $\lambda_R = 0.5$, and (b) the range of the repulsive interactions
           $\lambda_R$ at $\Gamma_A = 20$, showing the transition from finite bundles to infinite bundles at a fixed strength of the repulsive interactions $\Gamma_R = 5$ (and
           $\lambda_A = 0.3$) and $\rho^{\ast} = 1$. The bundle size has been scaled with respect to the in-plane
           radius of gyration of the polymers $\sqrt{L/\beta\epsilon}$, which plays the role of an effective polymer diameter.}
           \label{Fig3}
 \end{figure*}

We apply the instability condition to the special case of  Gaussian attraction that occurs very commonly in nature (hydrogen-bond, hydrophobic, cation-$\pi$, charge-bridging or
reversible cross-links ~\cite{silverstein1998simple,benetatos2013bundling}).
The local forms of the repulsive potential and the attractive potential are $\beta V_{R}(\Gamma_{R}, \lambda_{R}, \mathbf{r})
= \Gamma_{R}K_{0}(r/\lambda_{R})$ and $\beta V_{A}(\Gamma_{A}, \lambda_{A}, \mathbf{r}) = -\Gamma_{A}\exp(-r^2/\lambda_{A}^2)$ respectively (Appendix B in Supplementary \cite{Supplementary}).  
The distance $\mathbf{r}$ is dimensionless and has been scaled with respect to the in-plane radius of gyration $\sqrt{L/\beta\epsilon}$. 
$K_{0}(r)$ is the modified Bessel function of the second kind of order zero.
Figure \ref{Fig2}-(a) and \ref{Fig2}-(b) show the Fourier transform of the total potential $\beta V(\mathbf{k})$ and the corresponding stability function.
In our theory, the system forms bundles only when it satisfies the two conditions, (a) the Fourier potential is attractive $\beta V(\mathbf{k}) < 0$ and (b) 
the stability function $F(\mathbf{k})\geq 1$. When the total potential $\beta V(\mathbf{k})$ is repulsive, as it happens when the attractive potential is weaker 
than the repulsive potential, $\Gamma_{A} < \Gamma_{R}$, the condition (a) is violated and 
there is no instability. The system remains in a disordered liquid state. 
For moderately strong attraction $\Gamma_{A} \ge \Gamma_{R}$, (the curve (triangles) in Figure \ref{Fig2}-(b), the condition (b)) $F(\mathbf{k}) \ge 1$ is just satisfied
(at some finite  $\vert\mathbf{k}\vert = k_0 > 0$) in the attractive region of
the potential $\beta V(\mathbf{k}) < 0$ as shown in Fig \ref{Fig2}-(b). 
 In the position space, the system undergoes a transition to a phase with a modulation of the lateral (in-plane) areal density with a 
characteristic wavelength $2\pi/k_0$  which corresponds to the size of the finite bundles. 
For very strong attraction, denoted by the curve (squares) in Figure \ref{Fig2}-(a), the total potential is negative everywhere and $F(\mathbf{k}) \ge 1$ at the origin $\mathbf{k} = 0$. 
The entire system becomes unstable, the conditions (a) and (b) are trivially satisfied and macroscopic phase separation (infinite bundle) occurs.
The transition from finite bundles to an infinite bundle and the dependence of the size of the bundles on $\Gamma_A$ and $\lambda_R$ is shown in
Figure \ref{Fig3}. 

 According to the stability criterion of Eq. \ref{eq10}, polymers with infinite tension cannot undergo bundling. 
This is an artifact of the one-dimensional chain model that we use. The in-plane radius of gyration plays the role of an 
effective finite polymer diameter. Keeping this in mind, we see that our prediction for the bundle thickness as shown in Fig.
\ref{Fig3}  qualitatively agrees with experimental measurements. According to \cite{BorukhovPNAS}, 
the electrostatic screening length is about 1 nm whereas the filament thickness is about 8 nm. For these values,
Fig. \ref{Fig3} would give a bundle consisting of a few filaments in agreement with the findings of \cite{Claessens_Bausch}.

\begin{figure}[!htbp]
        \centering
           \onefigure[scale=0.3]{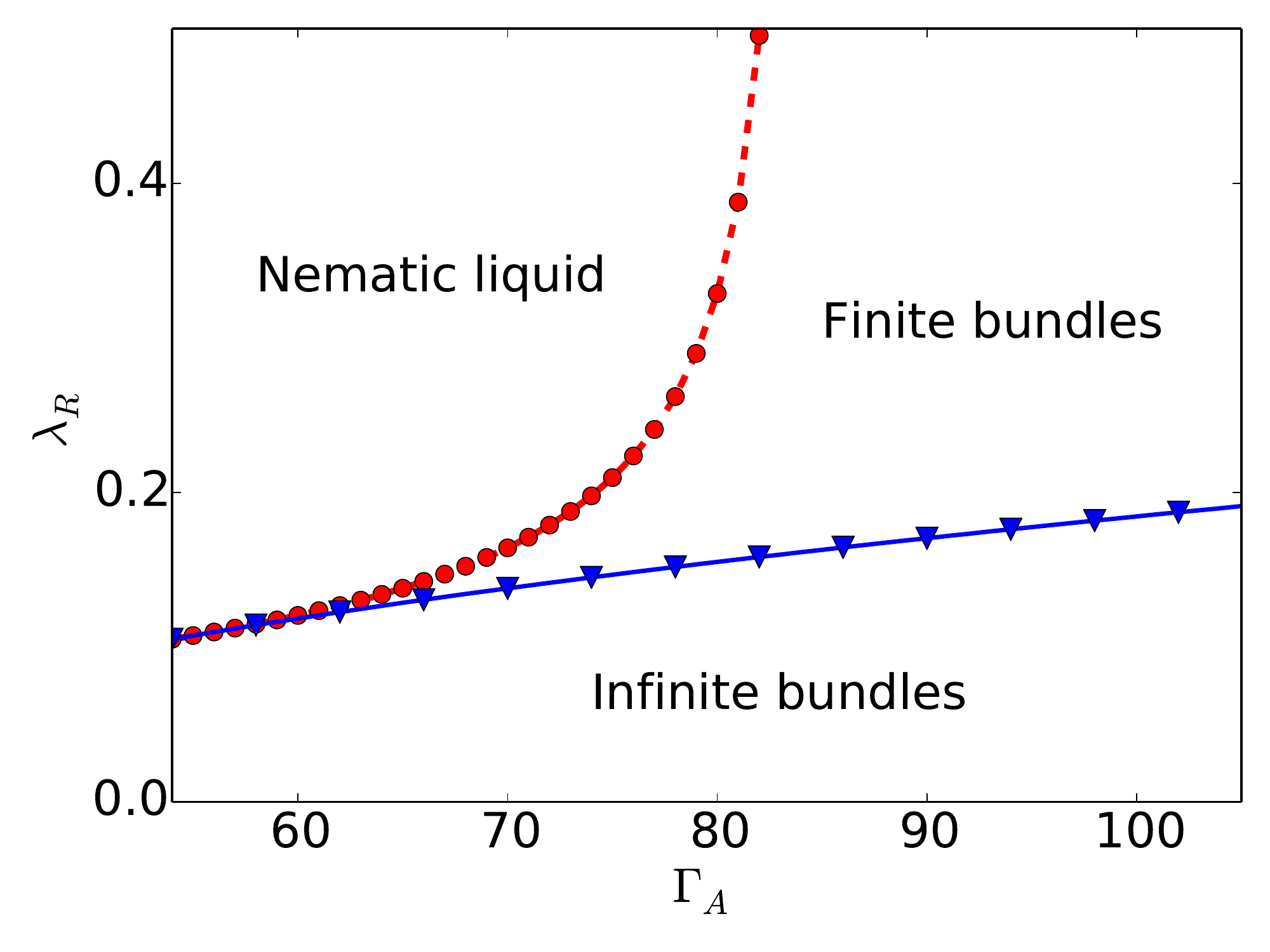}
           \caption{The stability diagram $\lambda_R$ vs $\Gamma_A$ predicted from  equation \eqref{eq10} showing the nematic fluid, finite bundle and infinite 
           bundle phases at $\Gamma_R = 10$, $\lambda_A = 0.1$ and $\rho^{\ast} = 1$. The $\Gamma$'s have been scaled with respect to $k_BT$ and the $\lambda$'s by the in-plane
           radius of gyration of the polymers $\sqrt{L/\beta\epsilon}$.}
           \label{Fig4}
 \end{figure}

The instability condition in equation \eqref{eq10} can be used to predict some useful concepts like the stability diagram.
The thermodynamics of the system depends on five parameters, namely,  $\rho^{\ast}$, $\lambda_A$, $\lambda_R$, $\Gamma_A$ and $\Gamma_R$. 
In Figure \ref{Fig4}, we show a stability diagram for two of these quantities $\lambda_R$ vs $\Gamma_A$ keeping the other quantities
fixed at $\rho^{\ast} = 1$, $\lambda_A = 0.1$ and $\Gamma_R = 10$. In the stability diagram, the dashed (circle) 
curve in Figure \ref{Fig4} is derived from the condition when the system just becomes unstable $F(\mathbf{k}) = 1$ at some
finite $\mathbf{k}$ in Figure \ref{Fig2}-(b) and 
 finite bundles just start to emerge. The solid (triangle) curve in Figure \ref{Fig4} corresponds to the condition when the system becomes unstable at the origin $F(\mathbf{k} = 0) = 1$
in Figure \ref{Fig2}-(b) causing macroscopic phase separation (infinite bundle). We can qualitatively use the stability diagram to explain the 
bundle formation in experiments by the addition of salts. In systems where the attraction is of non-electrostatic origin like 
cation-$\pi$ and hydrophobic interactions \cite{ruben1997alzheimer,perkin2006long,likechargecoacervate} and are on average weaker than the 
electrostatic repulsion, on increasing the salt concentration $\lambda_R$ decreases without significantly affecting 
the parameters $\Gamma_R$, $\Gamma_A$ and $\lambda_A$. Adding salt screens the electrostatic repulsion though the screening 
length $\lambda_R \propto 1/\sqrt{n}$ \cite{hansen1990theory}, where $n$ is the salt concentration. From Figure \ref{Fig4} we see that the system would move
from a disordered liquid state to a finite bundle state. In other experiments, salt causes the formation of very strong coordinate
bonds due to metal-ligand interactions \cite{xu2013mechanics}. Thus the salt strongly modifies the strength of attractive interactions $\Gamma_A$ but
not as much $\Gamma_R$ and $\lambda_R$. This would cause the system to directly transition from the liquid 
state to the infinite bundle state as shown in Figure \ref{Fig4}.

\textbf{Conclusion and Discussion} -
We investigate a general scheme for bundle formation in directed polyelectrolytes in the presence of a short-range attraction.
The control parameters are the strengths (determined {\it e.g.}, by the charges on the polymers, H-bonds, hydrophobicity, or cross-linkers)
and ranges (screening length) of the two competing interactions, and also the tension, the length, and the average areal density of the directed polymers.
Depending on the interplay of the control parameters, the nematic liquid phase which is translationally invariant in the 
transverse plane may become unstable. If the instability happens at a finite wave-number, a regular array of finite-sized bundles
emerges (microphase separation). If the instability happens at zero wave-number, macrophase separation occurs and a single 
infinite-sized bundle forms.

% There are several shortcomings in our analysis which need to be pointed out. 
 In our theoretical model, the parallel alignment is controlled
 by a mechanism different from the electrostatic repulsion and short range attraction which trigger the bundling transition. In many real systems,
 however, there is no aligning mechanism distinct from the two competing interactions. On the other hand, if the 
 bundled phase is a thermodynamic equilibrium phase, its structure and properties should not depend on the way it is prepared.  A general 
 theory which will describe bundling in stiff polymers with competing interactions without the assumption of pre-alignment is an open challenge 
 for future work. 
% In our model, we assume an attraction represented by a short-range pairwise potential.
%If this attraction is due to cross-links, it should be reversible and rather weak \cite{Kierfeld_Kuhne}.  
In our analysis, we neglect the
excluded-volume interaction.
As shown in Ref. \cite{benetatos2013bundling}, we could have added a delta-function excluded volume interaction (\`a la Edwards) which would 
simply shift the instability line without 
any further qualitative change. We have very crudely identified the characteristic  wavelength predicted from
the instability criterion with the size of the bundle. It should be noted that this is just a single length-scale for the polymer-dense and polymer-dilute regions of the microphase separation. In order to obtain the actual structure of the
bundled state, including the bundle width, one needs to go beyond the quadratic (spinodal) term in the Landau-Wilson theory and 
solve Eq. \eqref{eq9} self-consistently for $\Omega$.
 In our analysis, the parallel alignment of the polymers comes from their tension. Comparing with \cite{benetatos2013bundling}, 
 however, one can see that similar bundling would emerge in a brush of stiff semiflexible polymers (with large persistence length) 
 with shorter-range attraction and longer-range repulsion which are perpendicularly grafted on a fluid substrate (that is, the 
 grafting points free to move). This kind of bundling may be relevant to lamellipodia in the context of the actin gel-brush model 
 by Falcke {\it et al} \cite{Falcke_gel}.

\textit{Acknowledgments} -
%\label{Sec5}
PB  thanks the COSA Group at the NCSR Demokritos in Athens, Greece, for hospitality during part of this work. YSJ is 
supported by the
Ministry of Education, Science, and Technology (NRF-
2015R1D1A1A09061345, NRF-C1ABA001-2011-0029960) of
the National Research 
Foundation of Korea (NRF). SD thanks Yongjin Lee for helping us with the figures.

\bibliography{Bundle_EPL}
\bibliographystyle{eplbib}
% Create the reference section using BibTeX:
%\begin{thebibliography}{99} 
%\end{thebibliography}

\bigskip

\end{document}